\documentclass{iaus}
\usepackage{graphicx}
\usepackage{natbib}
\usepackage{bm}

\newcommand{\EQ}{\begin{equation}}
\newcommand{\EN}{\end{equation}}
\newcommand{\EQA}{\begin{eqnarray}}
\newcommand{\ENA}{\end{eqnarray}}

\newcommand{\Eq}[1]{Equation~(\ref{#1})}

\newcommand{\Fig}[1]{Figure~\ref{#1}}

\newcommand{\Tab}[1]{Table~\ref{#1}}

\newcommand{\bra}[1]{\langle #1\rangle}

{}
{}
{}

{}
{}
\newcommand{\meanEMF}{\overline{\mbox{\boldmath ${\cal E}$}}{}}{}
\newcommand{\tildeEMF}{\tilde{\mbox{\boldmath ${\cal E}$}}{}}{}
{}
{}
{}
{}
{}
\newcommand{\meanBB}{\overline{\mbox{\boldmath $B$}}{}}{}
\newcommand{\tildeBB}{\tilde{\mbox{\boldmath $B$}}{}}{}
{}
\newcommand{\meanFF}{\overline{\mbox{\boldmath $F$}}{}}{}
{}
{}
{}
{}
{}
\newcommand{\meanJJ}{\overline{\mbox{\boldmath $J$}}{}}{}

{}
{}
{}

{}

{}
{}

\newcommand{\zz}{\mbox{\boldmath $z$} {}}

\newcommand{\UU}{\mbox{\boldmath $U$} {}}

\newcommand{\bb}{\mbox{\boldmath $b$} {}}
\newcommand{\BB}{\mbox{\boldmath $B$} {}}

\newcommand{\jj}{\mbox{\boldmath $j$} {}}
\newcommand{\JJ}{\mbox{\boldmath $J$} {}}

\newcommand{\AAA}{\mbox{\boldmath $A$} {}}

\newcommand{\nab}{\mbox{\boldmath $\nabla$} {}}

\newcommand{\ii}{{\rm i}}

\newcommand{\DD}{{\rm D} {}}

\newcommand{\dd}{{\rm d} {}}

\def\Pm{\mbox{\rm Pr}_M}
\def\Rm{\mbox{\rm Re}_M}
\def\Rmc{\mbox{\rm Re}_{M,{\rm crit}}}

\def\kf{k_{\rm f}}

\def\Brms{B_{\rm rms}}
\def\Jrms{J_{\rm rms}}

\def\urms{u_{\rm rms}}

\def\etat{\eta_{\rm t}}

\def\Beq{B_{\rm eq}}
\def\epsK{\epsilon_{\rm K}}
\def\epsM{\epsilon_{\rm M}}
\newcommand{\yapj}[3]{ #1, \textit{ApJ,} \textit{#2}, #3}

\newcommand{\yapjs}[3]{ #1, \textit{ApJS,} \textit{#2}, #3}

\newcommand{\yan}[3]{ #1, \textit{Astron.\ Nachr.,} \textit{#2}, #3}

\newcommand{\yana}[3]{ #1, \textit{A\&A,} \textit{#2}, #3}

\newcommand{\ygafd}[3]{ #1, \textit{Geophys.\ Astrophys.\ Fluid Dyn.,} \textit{#2}, #3}

\newcommand{\yjetp}[3]{ #1, \textit{Sov.\ Phys.\ JETP,} \textit{#2}, #3}

\newcommand{\yaraa}[3]{ #1, \textit{ARA\&A,} \textit{#2}, #3}

\newcommand{\yprs}[3]{ #1, \textit{Proc.\ Roy.\ Soc.\ Lond.,} \textit{#2}, #3}

\newcommand{\yprl}[3]{ #1, \textit{Phys.\ Rev.\ Lett.,} \textit{#2}, #3}

\newcommand{\ymn}[3]{ #1, \textit{MNRAS,} \textit{#2}, #3}

\newcommand{\ypre}[3]{ #1, \textit{Phys.\ Rev.\ E,} \textit{#2}, #3}

\newcommand{\yjour}[4]{ #1, \textit{#2}, \textit{#3}, #4}

\title[Simulations of astrophysical dynamos]
{Simulations of astrophysical dynamos}

\author[A. Brandenburg]
{Axel Brandenburg}

\affiliation{NORDITA, AlbaNova University Center,
Roslagstullsbacken 23, SE-10691 Stockholm, Sweden;\\
Department of Astronomy, Stockholm University,
SE 10691 Stockholm, Sweden}

\pubyear{2010}
\volume{274}
\pagerange{119--126}
\jname{Advances in Plasma Astrophysics}
\editors{A. Bonanno, E. de Gouveia dal Pino \& A. Kosovichev, eds.}
\begin{document}

\maketitle

\begin{abstract}
Numerical aspects of dynamos in periodic domains are discussed.
Modifications of the solutions by numerically motivated alterations of the
equations are being reviewed using the examples of magnetic hyperdiffusion
and artificial diffusion when advancing the magnetic field in its
Euler potential representation.
The importance of using integral kernel formulations in mean-field
dynamo theory is emphasized in cases where the dynamo growth rate
becomes comparable with the inverse turnover time.
Finally, the significance of microscopic magnetic Prandtl number
in controlling the conversion from kinetic to magnetic
energy is highlighted.
\end{abstract}

\firstsection 
 \section{Introduction}

There are two important aspects connected with astrophysical dynamos
compared with dynamos on a bicycle.
Firstly, they are self-excited and do not require any permanent magnets.
Secondly, they are {\it homogeneous} in the sense that the medium is conducting
everywhere in the dynamo proper and there are no wires or insulators inside.
Self-excited dynamos were invented by the Danish inventor S{\o}ren Hjorth,
who received the patent for this discovery in 1854, some 12 years before
Samuel Alfred Varley, Ernst Werner von Siemens and Charles Wheatstone
announced such an invention independently of each other.
Von Siemens is known for having recognized its industrial importance
in producing the most powerful generators at the time, for which
he, in turn, received a patent in 1877.

The idea that homogeneous dynamos might work in the Sun, was first proposed
by \cite{Lar19} in a one-page paper.
However, some 14 years later, \cite{Cow33} showed that axisymmetric
dynamos cannot work in a body like the Sun.
At the time it was not clear whether this failure was genuine,
or whether it was critically connected with Cowling's assumption
of axisymmetry.
The suspicion that the third dimension might be critical was not
particularly emphasized when \cite{Lar34} tried to defend his early
suggestion with the words ``the self-exciting dynamo analogy is still,
so far as I know, the only foundation on which a gaseous body such as
the Sun could possess a magnetic field: so that if it is demolished
there could be no explanation of the Sun's magnetic field even remotely
in sight.''

The essential idea about the operation of the solar dynamo came
from \cite{Par55}, who developed the notion that cyclonic
events would tilt a toroidal field systematically in the poloidal
direction, closing thereby a critical step in the dynamo cycle.
While this concept is still valid today, it still required the
existence proof by \cite{Her58} that began to convince critics
that Cowling's antidynamo theorem does not extend to the general
case of three dimensions.

Nevertheless, subsequent progress in modeling the solar dynamo appears
to have been suspended until the foundations of a mean-field treatment
of the induction equation were developed by \cite{SKR66}.
In the following years, a large number of models were
computed covering mostly aspects of the solar dynamo
\citep{SK69a,Par70,Par70b,Par70c,Par71b,Par71d,Par71f}, but in some
cases also terrestrial dynamos \citep{SK69b,Par71c} and the galactic
dynamo \citep{Par71,Par71e,VR71,VR72}.
These developments provided a major boost to dynamo theory given that
until then work on the galactic dynamo, for example, focussed on aspects
concerning the small-scale magnetic field \citep{Par69}, but not the
global large-scale fields on the scale of the entire galaxy.
In fact, also regarding small-scale dynamos, there were important
developments made by \cite{Kaz68}, but they remained mostly unnoticed
in the West, even when the first direct simulations by \cite{MFP81}
demonstrated the operation of such a dynamo in some detail.
In fact, in some of these dynamos, the driving of the flow involved
helicity, but its role in helping the dynamo remained unconvincing,
because no large-scale field was produced.
We now understand that this was mainly because there was not enough scale
separation between the scale of the domain and the forcing scale, and that one
needs at least a ratio of 3 \citep{HBD04}.

Simulations in spherical geometry were much more readily able
to demonstrate the production of large-scale magnetic fields
\citep{Gil83,Gla85}, but even today these simulations produce
magnetic fields that propagate toward the poles \citep{KKBMT10}
and not toward the equator, as in the Sun.
We can only speculate about possible shortcomings
of efforts such as these that must ultimately be able to reproduce
the solar cycle.

Several important developments happened in the 1980s.
Firstly, it became broadly accepted that the magnetic field
inside the Sun might be in a fibril state \citep{Par82}, i.e.\
the filling factor is small and most of the field is concentrated
into thin flux tubes, as manifested by the magnetic field appearance
in the form of sunspots at the surface.
However, such tubes would be magnetically buoyant, and are expected to
rise to the surface on a time scale of some 50 days \citep{MI83,MI86}.
This time is short compared with the cycle time and might lead to
excessive magnetic flux losses, which then led to the proposal that the
magnetic field would instead be generated in the overshoot layer beneath
the convection zone.
This idea is still the basic picture today, although simulations
of convection generally produce magnetic fields that are distributed
over the entire convection zone.

Yet another important development in the 1980s was the proposal that
the $\alpha$ effect might actually be the sum of a kinetic and a
magnetic part and that the magnetic part can be estimated by
solving an evolution equation for the magnetic helicity density.
The importance of this development was obscured by the excitement
that the two evolution equations for poloidal and toroidal field,
supplemented by a third equation for the magnetic helicity density,
could produce chaos \citep{Ruz81}.
The connection to what was to come some 10--20 years later was not yet
understood at that point.
Simulations of helical MHD turbulence in a periodic domain demonstrated
that in a periodic domain the $\alpha$ effect might be quenched in an
$\Rm$-dependent fashion like
\EQ
\alpha={\alpha_0\over1+\Rm\meanBB^2/\Beq^2}.
\EN
If this were also true of astrophysical dynamos, $\alpha$ would be
negligibly small and would not be relevant for explaining the magnetic
field in these bodies.
Such quenching is therefore nowadays referred to as catastrophic quenching.
However, there is now mounting evidence that this type of
$\alpha$ quenching is a special case of a more general formula
\citep[see, e.g.,][]{B08}
\EQ
\alpha={\alpha_0+\Rm\left[
\eta_{\rm t}\mu_0\meanJJ\cdot\meanBB/B_{\rm eq}^2
-(\nab\cdot\meanFF_{\rm C})/2 k_{\rm f}^2B_{\rm eq}^2
-(\partial\alpha/\partial t)/2\eta_{\rm t} k_{\rm f}^2\right]
\over1+\Rm\meanBB^2/B_{\rm eq}^2},
\label{QuenchExtra2}
\EN
which comes from magnetic helicity conservation.
Note that in this equation there are 3 new terms that all scale with $\Rm$
and are therefore important.
Even in a closed or periodic domain, the first and third terms in squared
brackets contribute, the most promising way out of catastrophic quenching
is through magnetic helicity fluxes \citep{BF00,Klee00}.
These developments are still ongoing and we refer here to some
recent papers by \cite{SSSB06,BCC09,Can10}.

\section{Simulating dynamos}

\subsection{Roberts flow}

Much of the theoretical understanding of dynamos is being helped
by numerical simulations.
In fact, nowadays one of the simplest dynamos to simulate is the
Roberts flow dynamo.
A kinematic dynamo can be simulated by adopting a velocity field
of the form
\EQ
\UU=\nab\times\psi\zz+\kf\psi\zz,
\EN
where $\psi=(U_0/k_0)\cos k_0x\cos k_0y$.
In order to give some idea about the ease at which reasonably
accurate solutions can be obtained we give in \Tab{Tab1} the numerically
obtained critical values of the magnetic Reynolds number, $\Rm=\urms/\eta\kf$,
for a given resolution.

\begin{table}[h!]\caption{
Critical values of $\Rm$ for the Roberts flow dynamo
at low resolution (from $8^3$ to $32^3$ mesh points)
and different spatial order of the numerical scheme.
}\vspace{12pt}\centerline{\begin{tabular}{c|ccc}
           & \multicolumn{3}{|c}{--------- $\Rmc$ ---------} \\
Resolution & 2nd order & 6th order & 8th order \\
\hline
 $8^3$ & 5.23 & 5.15  & 5.16  \\
$16^3$ & 5.517& 5.514 & 5.518 \\
$32^3$ & 5.522& 5.522 & 5.521
\label{Tab1}\end{tabular}}\end{table}

Even turbulent dynamos are nowadays easy to simulate and meaningful
results have been obtained already at relatively low resolution, provided
the flow is helical \citep{B01}.
However, there are also examples where numerical aspects can have a
major effect on the outcome of such simulations.
In the following we discuss two examples: magnetic hyperdiffusion
and the use of Euler potential with artificial magnetic diffusion.

\begin{figure}[t!]
\centering
\includegraphics[width=\textwidth]{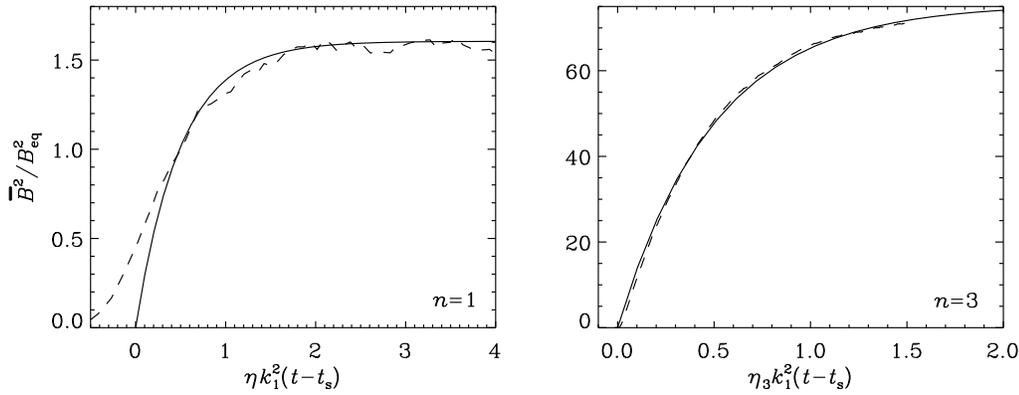}
\caption{
Evolution of large-scale magnetic field (dashed lines) for runs with regular
magnetic diffusion ($n=1$, left, $\urms/\eta k_1=45$)
and magnetic hyperdiffusion ($n=3$, right, $\urms/\eta_3 k_1=3200$),
at $16^3$ resolution and $\kf/k_1=3$,
compared with the prediction (solid lines) of \cite{BS02},
where $t_{\rm s}$ is the saturation time of the small-scale field.
}\label{pcomp}
\end{figure}

\subsection{Magnetic hyperdiffusion in helicity-driven dynamos}

Dynamos work by maintaining the magnetic field against Ohmic decay
via magnetic induction.
It is then not surprising that the result can be sensitive to the
numerical treatment of magnetic diffusion.
One example is the consideration of magnetic hyperdiffusion of the form
\EQ
\eta\JJ\to(-1)^{n-1}\nu_n\nabla^{2n-2}\JJ
\EN
instead of the regular $\eta\JJ$ term.
In helical dynamos in closed or periodic domains the saturation time
is given by $\eta k_1^2\Delta t=1$, but with hyperdiffusion this
condition becomes $\eta_n k_1^{2n}\Delta t=1$; see \Fig{pcomp}.
This is exactly what one expects from a scheme like hyperdiffusion
that enhances the effective magnetic Reynolds number.
However, another important modification is that the saturation
amplitude increases from \citep{BS02}
\EQ
{\meanBB^2\over\Beq^2}\approx{\kf\over k_1}
\quad\mbox{to}\quad
{\meanBB^2\over\Beq^2}\approx\left({\kf\over k_1}\right)^{2n-1}.
\label{BS02eqn}
\EN
This is a caveat that is important to keep in mind when employing
magnetic hyperdiffusion for astrophysical simulations.
This has a major effect on the saturation of large-scale magnetic
fields that is at first glance surprising.
However, once one realizes that the saturation in a periodic domain is
governed by the magnetic helicity equation applied to the steady state,
i.e.\ by
\EQ
{\dd\over\dd t}\bra{\AAA\cdot\BB}=-2\eta\bra{\JJ\cdot\BB},
\EN
where we split the right-hand side into contributions from large- and
small-scale fields, i.e.\ $\BB=\meanBB+\bb$ and $\JJ=\meanJJ+\jj$, so that
$\bra{\JJ\cdot\BB}=\bra{\meanJJ\cdot\meanBB}+\bra{\jj\cdot\bb}$,
we have for fully helical large- and small-scale fields in the steady state,
\EQ
0=-2\eta\bra{\meanJJ\cdot\meanBB}-2\eta\bra{\jj\cdot\bb}
\approx\pm2\eta\left(k_1\bra{\meanBB^2}-\kf\bra{\bb^2}\right),
\label{balance}
\EN
with $\bra{\bb^2}\approx\Beq^2$, it becomes clear that the use of
magnetic hyperdiffusion picks up the $k_1$ and $\kf$ factors at
correspondingly higher powers, leading thus to \Eq{BS02eqn}.
The upper and lower signs of the term on the right-hand side of
\Eq{balance} apply to small-scale forcings with positive and negative
helicity, respectively.

\subsection{MHD with Euler potentials}

\begin{figure}[t!]
\centering
\includegraphics[width=.61\columnwidth]{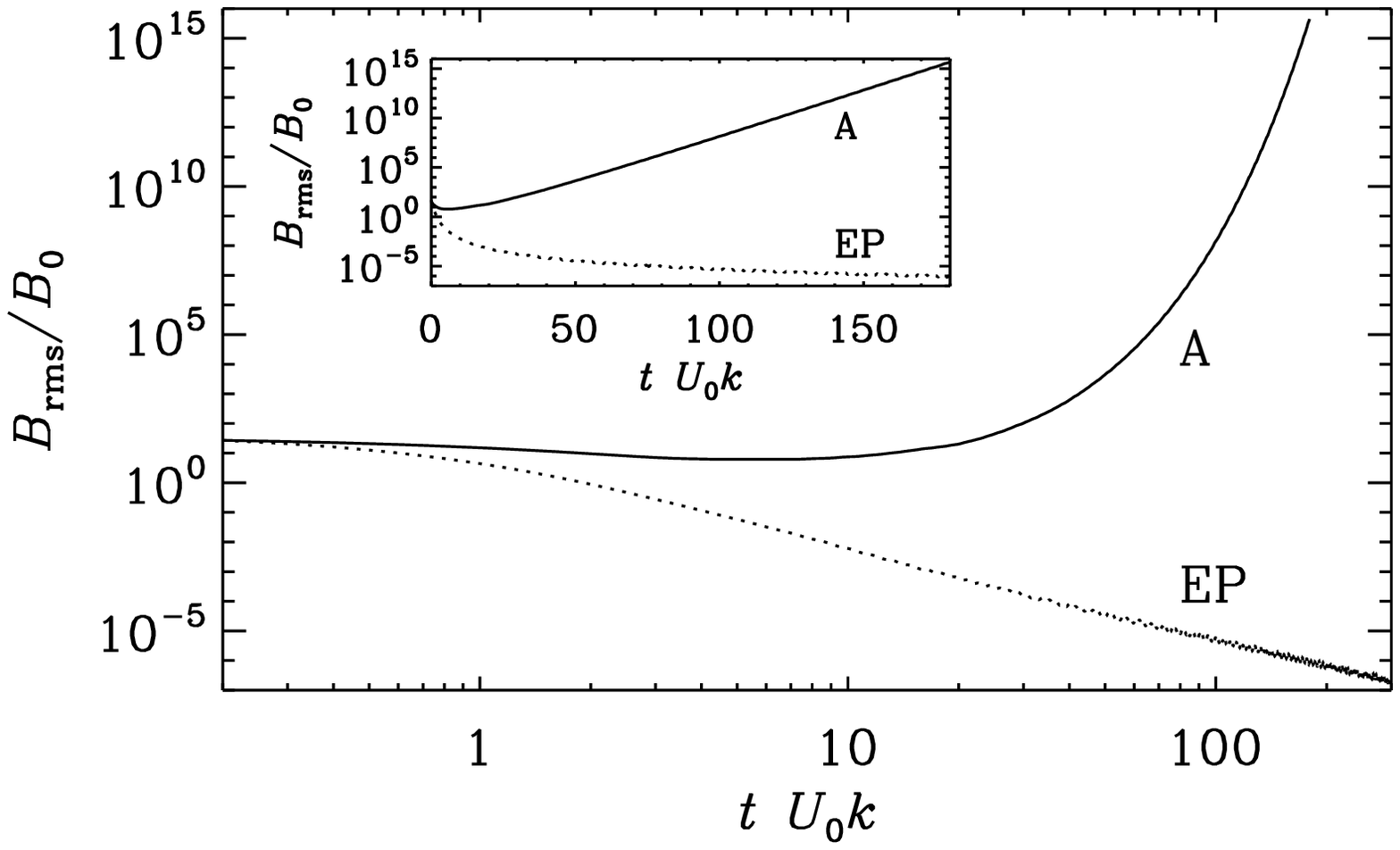}
\includegraphics[width=.37\columnwidth]{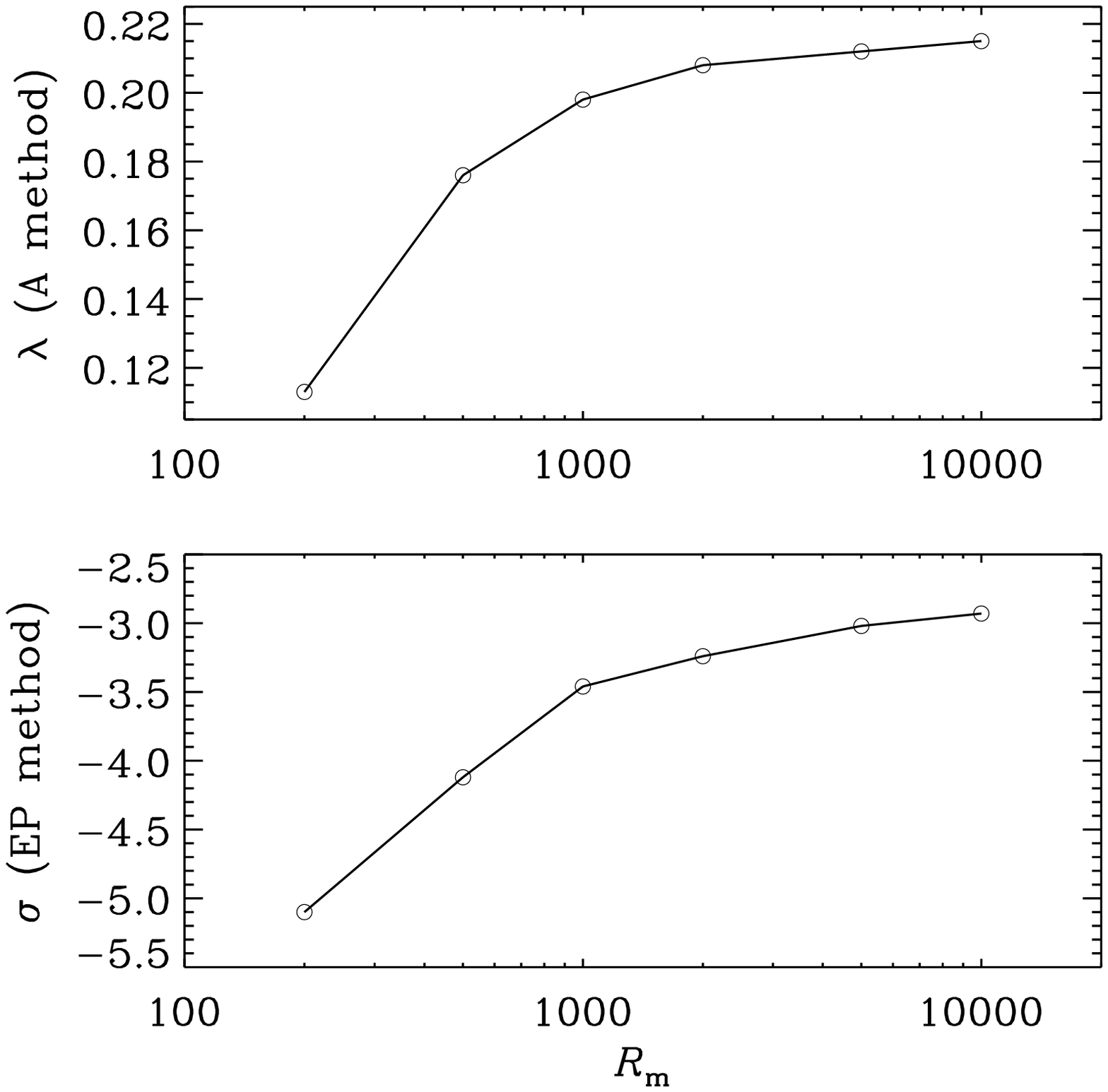}
\caption{
Left: Comparison of the evolution of $B_{\rm rms}$ for the modified
Galloway--Proctor flow with point-wise zero helicity
for methods~A and EP using $256^3$ meshpoints and $\Rm=10^4$.
Note the power law scaling for the EP method and the exponential
scaling for the A method.
Right: $\Rm$ dependence of the exponents $\lambda$ and $\sigma$
characterizing the evolution of $\Brms\sim e^{\lambda t}$ for the A method
and $\Brms\sim t^\sigma$ for the EP method for the modified
Galloway--Proctor flow with point-wise zero helicity
for methods~A and EP using $256^3$ meshpoints.
Adapted from \cite{B10}.
}\label{pt_HCK95_256d}
\end{figure}

Until recently, the use of Euler potentials (EP) has been a popular choice
for solving the MHD equations numerically using Lagrangian methods
\citep{PB07,RP07}.
The representation of $\BB$ in terms of EP, $\alpha$ and $\beta$, as
\EQ
\BB=\nab\alpha\times\nab\beta
\EN
is a nonlinear one, which solves the induction equation
in the case $\eta=0$, $\partial\BB/\partial t=\nab\times(\UU\times\BB)$,
provided $\DD\alpha/\DD t=\DD\beta/\DD t=0$.
The problem is that the magnetic field tends to develop sharp structures
that will not be properly resolved by the numerical scheme.
The hope has been that the overall properties of the magnetic field
at larger scales would then still be approximately correct.
However, this turns out not to be the case.
And, more importantly, as one increases the resolution, the solution
converges, but it is simply the wrong solution.
This has been demonstrated in detail in a separate paper \citep{B10}
where, among other cases, solutions of the Galloway--Proctor fast
dynamo flow were considered.
In \Fig{pt_HCK95_256d} we show that the solution of the induction equation
using EP leads to algebraically decaying solutions of the form
\EQ
\Brms\sim t^\sigma\quad\mbox{(EP method)},
\EN
where, with increasing resolution, $\sigma$ converges
toward a value around $-3$, while with the usual vector potential
method (A method) one solves $\partial\AAA/\partial t=\UU\times\BB-\eta\JJ$
with $\BB=\nab\times\AAA$ and $\JJ=\nab\times\BB$, and finds instead
\EQ
\Brms\sim e^{\lambda t}\quad\mbox{(A method)},
\EN
where $\lambda$ also converges, and its value is about 0.22 for $\Rm=10^4$;
see \Fig{pt_HCK95_256d}.

It is quite remarkable that by changing the properties of the solution
at small scales only slightly, one can produce rather dramatic effects.
This includes cases of magnetic hyperdiffusivity, where the large-scale
field amplitude can be quite different, albeit in agreement with the
theory applied to the hyperdiffusive case \citep{BS02}.
Another example is that of artificial diffusion in solutions for the
EP, where the obtained results bear no resemblance with those obtained
using the A method.

\subsection{Quantitative comparison between simulations and mean-field theory}

Mean-field theory has the potential of being a quantitatively accurate and
hence predictive theory.
In order to establish this in particular cases, it is important to
consider as many contact points between theory and simulations as possible.
One thing we can do is to determine $\alpha$ effect, turbulent diffusion,
and other effects from simulations, and to compare the thus calibrated
mean-field model with simulations.
This provides an important resource for ideas of what one might have
been missing in various contexts.
Here we just mention the case of the Roberts flow, for which $\alpha_{ij}$
and $\eta_{ij}$ have been determined using the test-field methods \citep{Sch07}.
One might then expect that the growth rate obtained from the underlying
dissipation rate,
\EQ
\lambda=\alpha k-(\etat+\eta)k^2
\EN
should agree with the value obtained from the direct calculation.
This is however only the case for $\lambda=0$, but not for $\lambda\neq0$.
However, it would be a mistake to assume that there is something
wrong with the test-field methods.
Instead, what is wrong here is just the assumption of homogeneity and
stationarity.
Obviously, when $\lambda\neq0$, the solution is exponentially growing
or decaying like $e^{\lambda t}$, so it is clearly not steady!

\begin{figure}[t!]
\centering\includegraphics[width=\columnwidth]{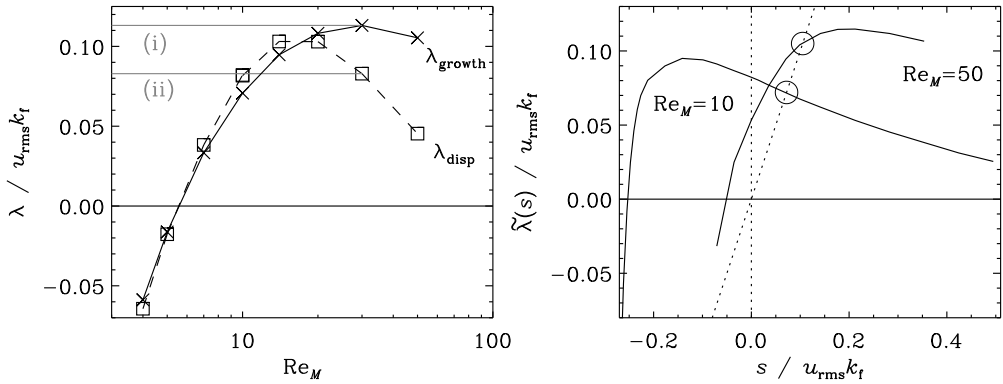}
\caption{
Left:
$\Rm$ dependence of the growth rate for the Roberts flow as obtained
from a direct calculation ($\lambda_{\rm growth}$) compared with the
result of the dispersion relation,
$\lambda_{\rm disp}=\alpha k_z-(\eta+\etat)k_z^2$, using a
cubic domain of size $L^3$, where $k_1=2\pi/L$ and $\kf=\sqrt{2}k_1$.
For this range of $\Rm$, the most unstable mode is the largest one
that fits in the box ($k_z=k_1$).
The two horizontal lines in gray mark the values of $\lambda_{\rm growth}$
and $\lambda_{\rm disp}$ at $\Rm=30$, denoted by (i) and (ii), respectively.
Right:
Laplace-transformed effective growth rate,
$\tilde\lambda(s)=\tilde\alpha(s)k-[\eta+\tilde\etat(s)]k^2$,
for the Roberts flow with $\Rm=10$ and $50$.
Note the different signs of the slope at the intersection with the diagonal
(denoted by circles).
Adapted from \cite{HB09}.
}\label{pscan_Rm}\end{figure}

When the assumption of steadiness is no longer satisfied, one has to return
to the underlying integral relation between $\alpha$ and the mean field,
i.e.\ $\alpha(z,t)\meanBB(z,t)$ has to be replaced by a convolution, so
\EQ
\alpha(z,t)\meanBB(z,t)\to\int\alpha(z-z',t-t')\meanBB(z',t')\,\dd z'\,\dd t',
\EN
and similarly for the magnetic diffusivity.
One way of dealing with this complication is to note that the convolution
in real space corresponds to a multiplication in Fourier space.
In other words, we can write
\EQ
\tildeEMF(k,\omega)=\tilde\alpha(k,\omega)\tildeBB(k,\omega),
\EN
where
$\tildeEMF(k,\omega)=\int\meanEMF(z,t)\,e^{-i(kz-\omega t)} \,\dd z\,\dd t$
is the Fourier-transformation of $\meanEMF(z,t)$ (and likewise for the
other fields).
Of course, the value of $\omega$ that is of interest is $\omega=\ii\lambda$,
where $\lambda$ is the then self-consistently obtained growth rate.
This has been described in detail by \cite{HB09}, who motivated
their study using the example of the Roberts flow, where the
discrepancy between the numerical solution and that obtained for
$\lambda=0$ is shown.
A reasonable fit to their data is $\alpha(k,\omega)=1/(1-\ii\omega\tau)$
for $k\to0$.

\section{Low magnetic Prandtl number and application to accretion discs}

In many astrophysical bodies the magnetic Prandtl number, $\Pm=\nu/\eta$,
is either large or small, but not around unity.
Again, from a numerical point of view, it is surprising that the
ratio $\nu/\eta$ is important even though
\EQ
\nu\to0,\quad\eta\to0.
\EN
In many numerical treatments it is implicitly assumed that the exact
values of $\nu$ and $\eta$ do not explicitly matter, because it should
not matter how long each of the two turbulent cascades is.
This should indeed be true provided the dynamics we are interested in
takes place entirely on the large scales.
But for helical large-scale dynamos, this is evidently not the case,
because in the kinematic regime, the magnetic energy spectrum
follows the Kazantsev $k^{3/2}$ spectrum and reaches a peak at
the resistive scale, corresponding to the wavenumber $k_\eta=(\Jrms/\eta)$.
However, this applies first of all only to the kinematic regime and,
seemingly, it is relevant only for the small-scale dynamo regime.
This has been discussed recently in connection with the contrasting
case of large-scale dynamos, where the excitation conditions are
not affected by the value of $\Pm$ \citep{B09,B11}.
Nevertheless, even in this case the dissipation rates of kinetic
and magnetic energies, $\epsK$ and $\epsM$, respectively, do depend
on the value of $\Pm$.
Although it seems fairly clear that the ratio $\epsK/\epsM$ increases
with $\Pm$ in power law fashion proportional to $\Pm^n$, the exponent $n$
is not well constrained.
Earlier work covering the range $10^{-3}\leq\Pm\leq1$ at $512^3$
resolution suggested $n=1/2$, although additional data covering
also the range $1<\Pm\leq10^3$ given an exponent closer to $n=0.6$ or
even $n=2/3$.
There is at present no theory for the value of the exponent $n$.

Applying these findings to accretion discs, we should first recall
that the work of \cite{LL07} suggested that the onset of the
magneto-rotational instability shows a strong dependence on $\Pm$.
However, one should expect that when large-scale dynamo action is
possible, this condition may change and the onset would then be
independent of $\Pm$.
This is indeed what \cite{KK10} find.
The question is of course what is the relevant large-scale
dynamo mechanism in this case.
One proposal is the incoherent $\alpha$--shear dynamo, which works
through constructive amplification of the mean field in the direction
of mean shear.
Yet another possible mechanism is the shear--current dynamo
\citep{RK03}, although this result has not yet been confirmed.

\section{Conclusions}

Simulating simple dynamos on the computer is nowadays quite simple.
Nevertheless, we have seen here examples that illustrate that things
can also go quite ``wrong''.
In the case of magnetic hyperdiffusion, it is clear what happens \citep{BS02},
so that magnetic hyperdiffusion can also be used to ones advantage,
as was demonstrated in \cite{BDS02}.
However, in the case of Euler potentials it is not clear what happens
and whether this method can be used to simulate even the ideal MHD equations,
given that each numerical scheme will introduce some type of diffusion.
In this short review, we have also attempted to clarify why numerical
calculations of $\alpha$ effect and turbulent diffusion using the
standard test-field method \citep{Sch07} would yield values that can
only reproduce a correct growth rate in the case of vanishing growth.
In all other case, a representation in terms of integral kernels
has to be used.
Finally, we have discussed some effects of using magnetic Prandtl numbers
that are different from unity.
It turns out that in the steady state, the rate of transfer from kinetic
to magnetic energy depends on the value of $\Pm$.
This is somewhat unexpected, because the onset condition for dynamo action
does not depend on $\Pm$ \citep{B09}, and yet the actual efficiency of
the dynamo, as characterized by the work done against the Lorentz force,
$-\bra{\UU\cdot(\JJ\times\BB)}$, does depend on $\Pm$ and is proportional
to $\Pm^{-n}$ (with $n$ between 1/2 and 2/3) for large values of $\Pm$.

Understanding the limits of numerical simulations is just as important
as appreciating its powers.
As the example with the problem with mean-field and simulated
growth rates shows, understanding the initial mismatch can be the
key to a more advanced and more accurate theory that will ultimately
be needed when describing some of the yet unexplained properties of
astrophysical dynamos.

\end{document}